\documentclass[11pt]{article}
\usepackage{moriond,graphicx,amsmath,bm}
 
\begin{document}
\vspace*{4cm}
\title{QCD THEORY AT THE XL RENCONTRE DE MORIOND:\\
FISH EYES AND PHYSICS}

\author{ D.E.~SOPER }

\address{Department of Physics, University of Oregon,\\
Eugene, OR 97403, USA}

\maketitle

\vskip -5 cm
\centerline{\includegraphics[width = 3.7 cm]{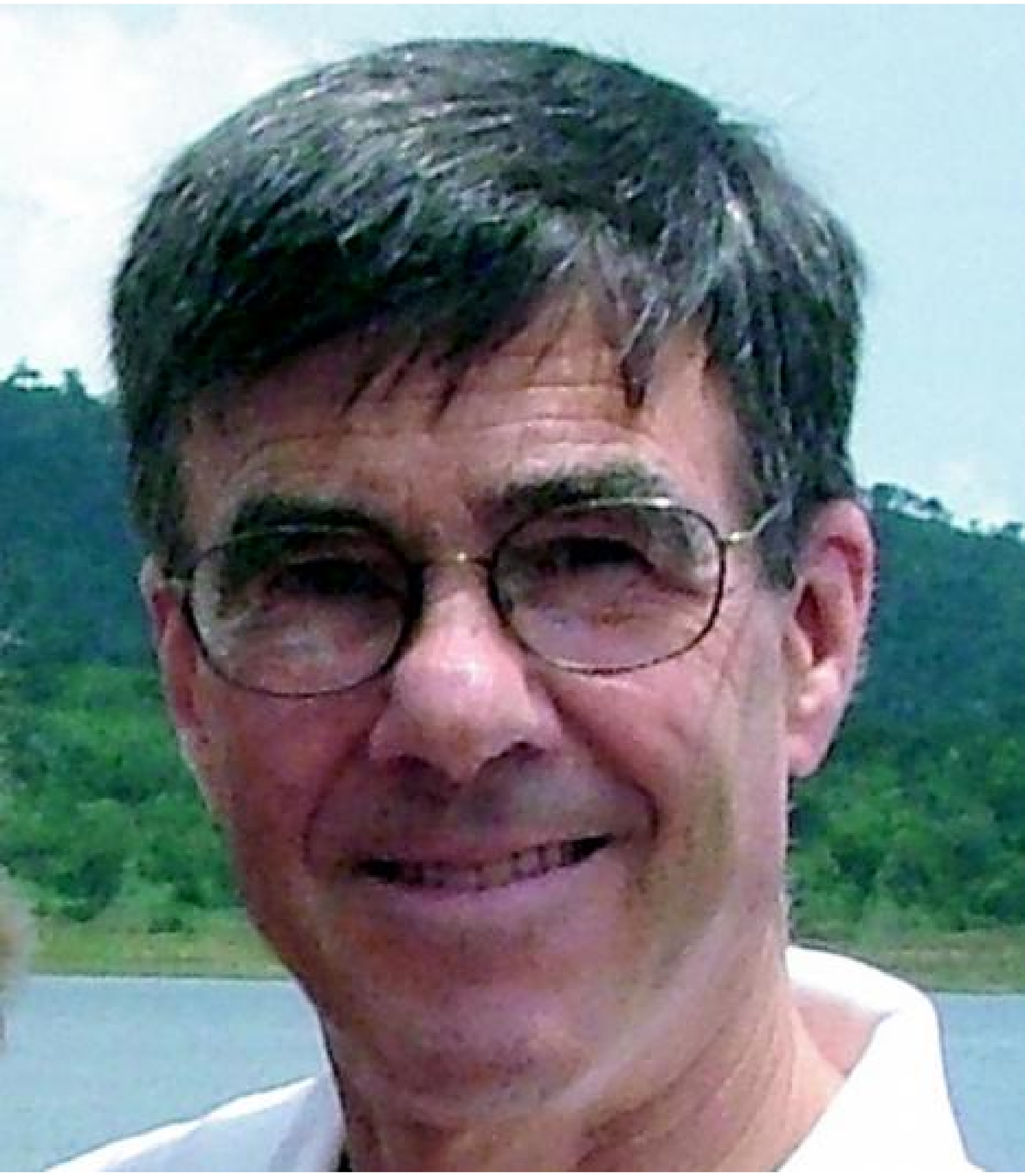} }
\vskip 0.5 cm

\abstracts{
I review a selection of the talks at the QCD Session of the XL
Rencontre de Moriond, talks either by theorists or else of special
theoretical interest. I use the talks to provide some assessment of
where we stand with respect to the problems and opportunities facing QCD
theory. }

\section{Introduction}

The theory of the strong interactions is well understood in one sense: we
are confident that the simple lagrangian of Quantum Chromodynamics (QCD)
provides the correct description of the strong interactions. After all, it
has passed many experimental tests that could have proven it wrong. Yet we
face challenges in applying that theory. Predictions for bound state
masses are difficult, as are predictions for decays of mesons containing
a heavy quark, where we need the QCD part of the theory in order to use
experimental results to get at the electroweak part. There appears to be
a deep simplicity in the behavior of densely packed gluons, as probed at
``small x,'' but this behavior is not well understood. Interesting
regularities in the development of the final state in heavy ion
collisions have been observed, but have not been susceptible to a fully
satisfactory interpretation in terms of quark-gluon interactions.
Finally, we seek more powerful tools for using the theory in the context
of high $p_T$ reactions at the upcoming Large Hadron Collider (LHC).
Surely, we imagine, new physics signals will be found there, but
understanding what those signals mean in terms of new particles or forces
-- maybe even new dimensions of space-time -- will not be easy.
We will want to use all of the theoretical methods we can muster.

These issues and more came up at the conference. In this talk, I review a
selection of the talks that were either presented by theorists or were of
special theoretical interest. I use the talks to provide some
assessment of where we stand with respect to the problems and
opportunities facing QCD theory. You have to read to the end to find out
about the fish part.

\section{b-quark production}

\begin{figure}[htb]
\centerline{
\includegraphics[width = 6 cm]{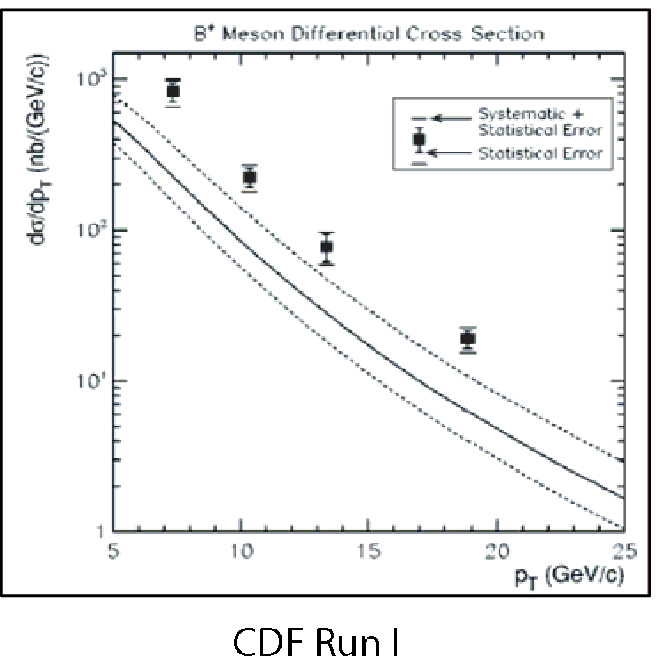}
\includegraphics[width = 7 cm]{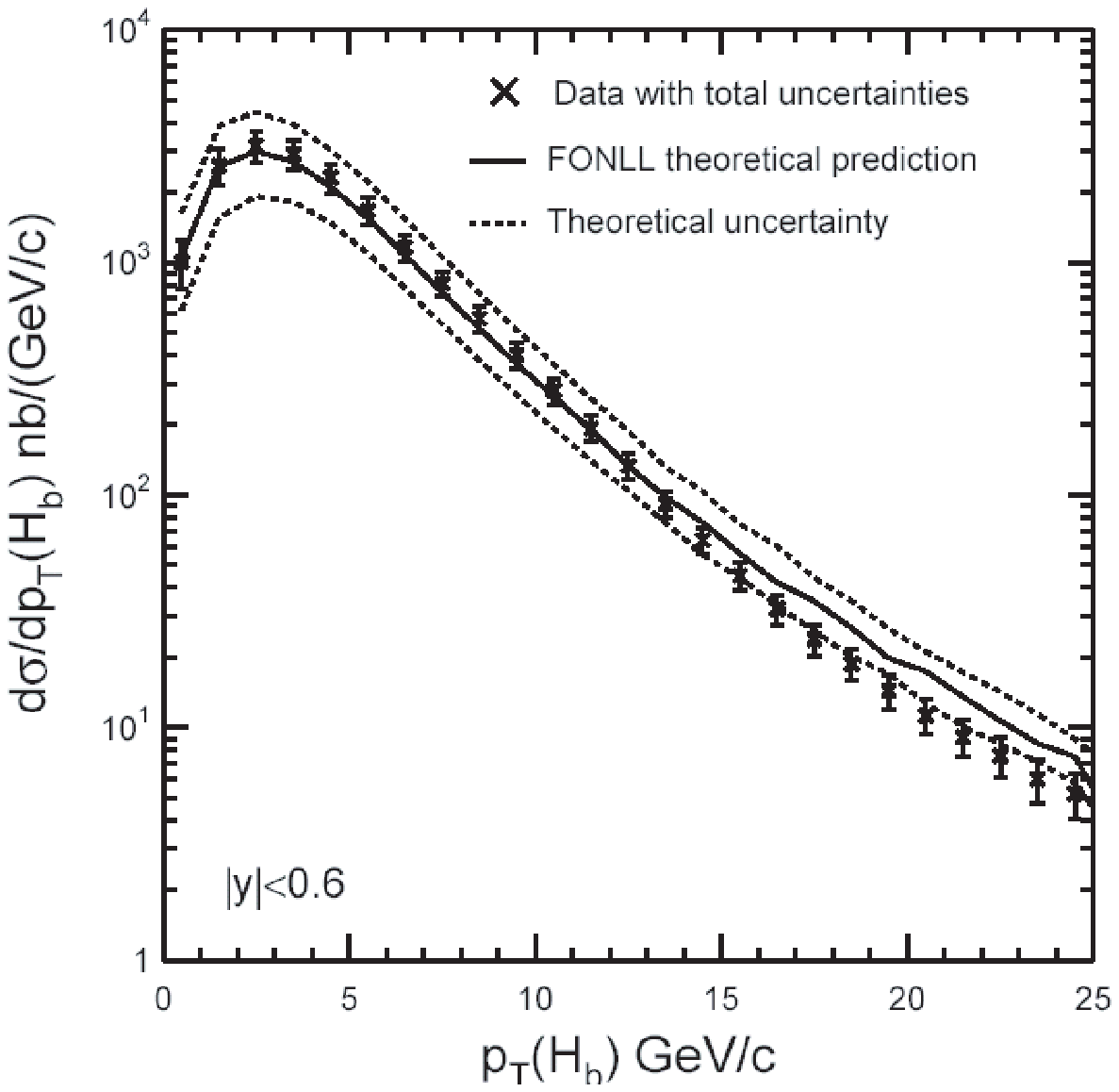}}
\caption{On the left, CDF data from about 1995 for $B^+$ meson production
cross section versus $P_T$ from the talk of
I.~Baliev.\protect\cite{baliev} On the right, current CDF data for B
hadron production from the talk of M.~d'Onofrio.\protect\cite{dOnofrio}}
\label{bquark}
\end{figure}

In the bad old days, around Moriond XXX, the data on b-quark production
indicated a problem, as illustrated on the left in Fig.~\ref{bquark}. The
results varied somewhat according to year, what was measured, and how the
measurement was performed, but always indicated the observation of more
$b$-quarks than theory predicted. One could hide behind the excuse that
the $P_T$ values were never more than about 20 GeV and the mass of the
$b$-quark itself is only 5 GeV, so that the effective $\alpha_s$ was not
so small and the effectiveness of the perturbative QCD approach could be
doubted. Nevertheless, one was left with the suspicion that the $b$ quark
had some kind of anomalous behavior that lay beyond the Standard Model.
In the last few years, the situation has been changing. At this
conference, M.~d'Onofrio presented the latest CDF and D0 data on this
subject.\cite{dOnofrio} I present one of the results on the right in
Fig.~\ref{bquark}. We see that the theory errors are still rather
substantial, reflecting the difficulties mentioned above. The
experimental errors are improved. And now data and theory agree.

There was some discussion at the conference of what had changed. First,
we have better theory that sums logs of $p_T/m_b$. Second, we have better
parton distributions and accompanying parton distributions. (See
d'Onofrio's talk.) Third, we measure a quantity whose definition does not
involve too much theory, for example we measure cross sections for
B-hadron production instead of b-quark production as was sometimes done in
the past. A recent trend has been to measure the cross section for jets
containing a b-quark.  Examples of this were shown, although so far the
NLO theory calculations are lacking and summing logarithms of $p_T/m_b$
will be needed to supplement a purely NLO calculation.

\section{b-quark decays}

There were several talks concerning the theory of b-quark decays. These
talks concerned perturbative calculations of the effective lagrangian,
lattice gauge theory calculations of decay matrix elements, and
theoretical analyses of the decay spectrum in inclusive decays.

\begin{figure}[htb]
\centerline{\includegraphics[width = 6 cm]{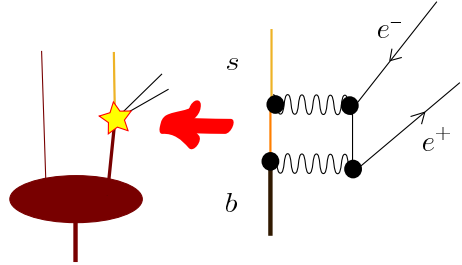}
}
\caption{The effective Lagrangian for b-quark decay. The loop diagram
illustrated creates an effective point interaction for $b \to s e^+e^-$
since the momentum in the loop is of order $M_W$, which is much larger
than $m_b$. This point interaction can then be used in a calculation of B
meson decay.}
\label{Leff}
\end{figure}

The issue of the effective lagrangian is illustrated in Fig.~\ref{Leff}.
Loop diagrams create an effective lagrangian of the form
\begin{equation}
{\cal L} = \sum C_i {\cal O}_i
\label{effL}
\end{equation}
for the process $b \to s e^+ e^-$ (for example).  The ${\cal
O}_i$ are operators such as $\bar s(x) \gamma^\mu b(x)\,
\bar e(x) \gamma_\mu e(x)$. The coefficients $C_i$ are to be calculated
from the loop diagrams. My example diagram is too simple; what we need
now are two loop diagrams. T.~Huber reported his calculation of the $C_i$
for $b \to s e^+ e^-$ from two loop diagrams in the Standard
Model.\cite{Huber}  S.~Schilling described a calculation
of the $C_i$ for $b \to s e^+ e^-$ in the case of the two higgs doublet
model.\cite{Schilling}

\begin{figure}[htb]
\centerline{\includegraphics[width = 3 cm]{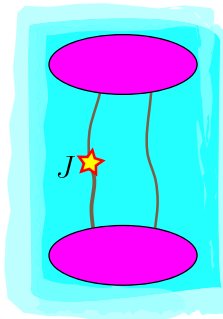} }
\caption{Calculation of the matrix element for a heavy meson decay using
lattice gauge theory. Here $J$ stands for the quark part of one of the
operators ${\cal O}_i$ in Eq.~(\ref{effL}). The operator changes a heavy
quark to a light quark and we need its matrix element between meson
states.}
\label{latticeME}
\end{figure}

If we wish to calculate a completely inclusive decay rate, not much is
needed beyond the effective lagrangian for the decay. But if we want an
exclusive decay rate, we need a matrix element of the effective
lagrangian between initial and final states, as illustrated in
Fig.~\ref{latticeME}. M.~Okamoto presented results for the matrix
elements of appropriate weak decay operators between an initial heavy
meson state and a final light meson state.\cite{Okamoto} The results are
needed for a range of momentum transfers $q$ to the quark system. With
appropriate limiting procedures, such matrix elements can be evaluated in
lattice QCD. In the calculations described by Okamoto, enough results can
be obtained to extract the complete CKM matrix from the corresponding
experimental results.

\begin{figure}[htb]
\centerline{\includegraphics[width = 6 cm]{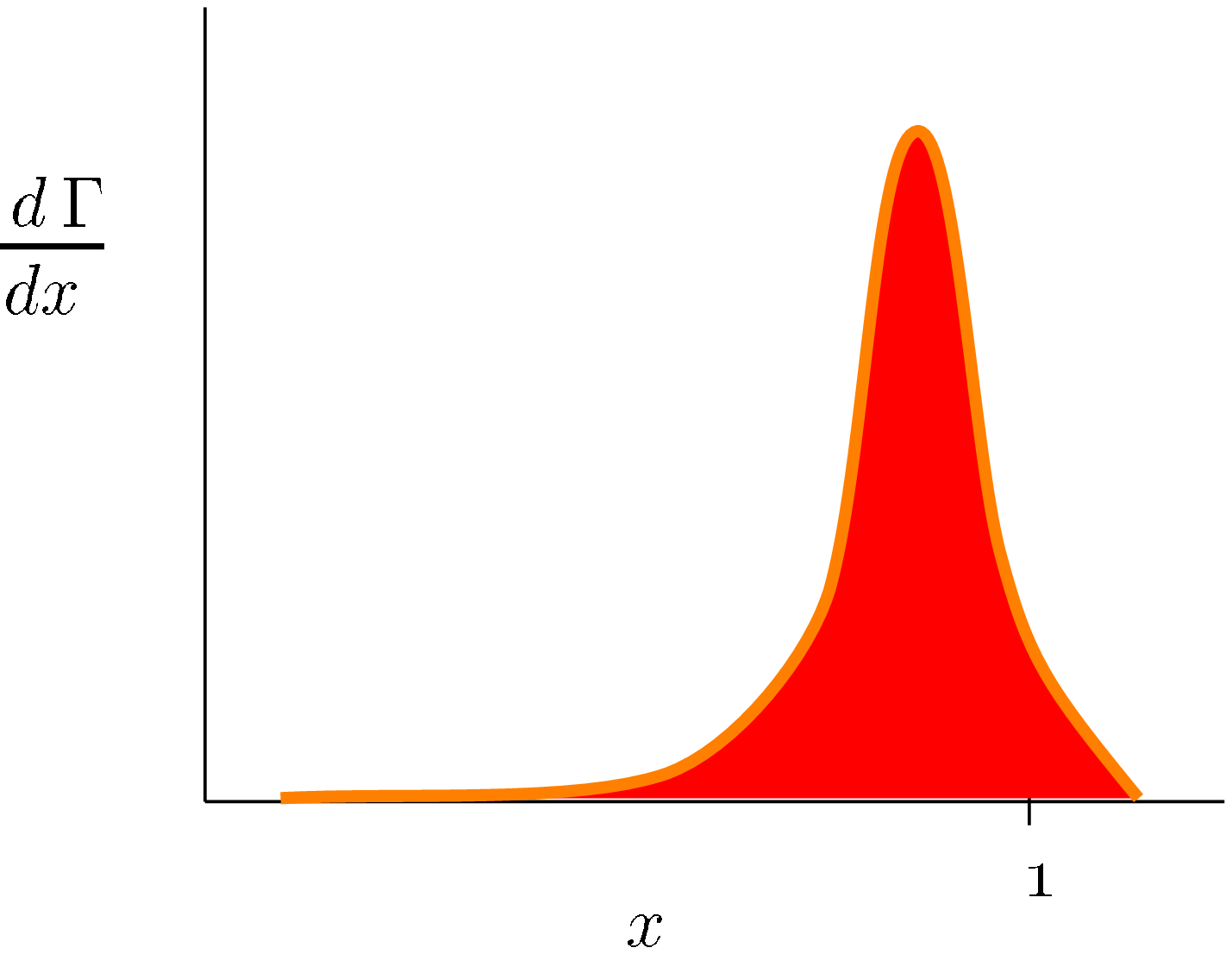} \hskip 1 cm
\includegraphics[width = 6 cm]{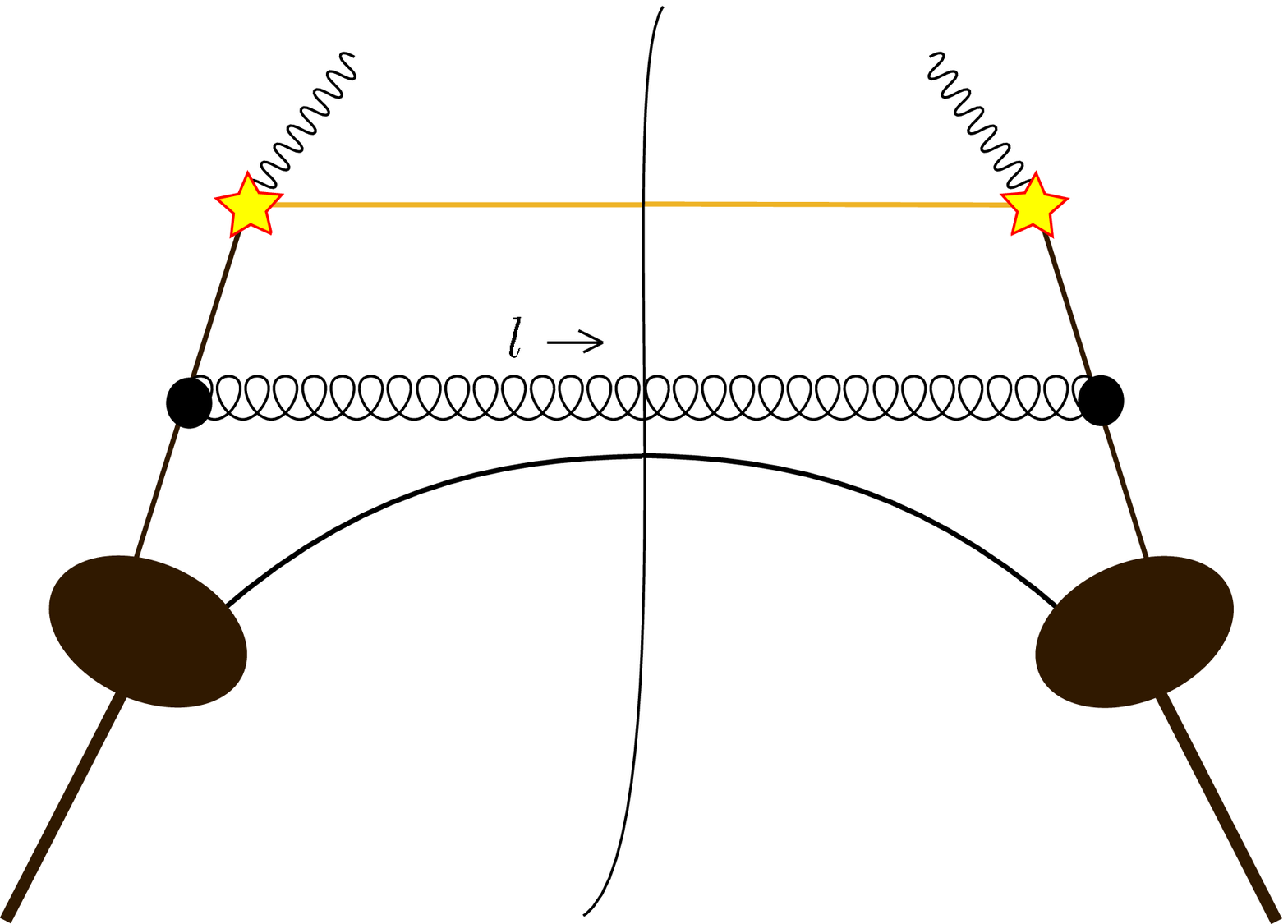}}
\caption{Left: sketch of the behavior of the differential decay width for 
$B \to X_s\gamma$ as a function of $x = 2 E_\gamma/M_b$. Right: a cut
Feynman diagram illustrating a contribution to $d\varGamma/dx$.}
\label{btosgammavsx}
\end{figure}

I turn now to the decay spectrum in inclusive decays. Consider, in
particular, the decay $B \to X_s\gamma$, where $X_s$ indicates any state
that has an $s$-quark in it. Let $x = 2 E_\gamma/M_b$. From a theoretical
point of view, by far the simplest cross section to calculate is the
inclusive cross section $\int dx\, d\varGamma/dx$. However, the
experimental acceptance typically does not include all $x$. For this
reason, we need theory for $d\varGamma/dx$ as a function of $x$. In the
simplest approximation, the $b$ quark in a $B$ meson carries all of the
four-momentum of the meson and decays into a two body state $s\gamma$.
Then $d\varGamma/dx = \delta(1-x)$. In a more realistic picture, we
expect the photon spectrum to be spread out, as illustrated in the
left-hand part of Fig.~\ref{btosgammavsx}. If one calculates Feynman
diagrams like the simple one in the right-hand part of
Fig.~\ref{btosgammavsx}, one gets contributions that are singular at $x
\to 1$ (before smearing by the wave function). In the region $(1-x) \ll
1$, one can sum a series containing more and more powers of $\log(1-x)$.
However, this series does not converge well. E.~Gardi \cite{Gardi}
reported on this. He blamed the bad behavior of the perturbative series
on an infrared renormalon, which is associated with a factor of $n!$ at
$n$th order in perturbation theory. The simple way to understand this is
that diagrams like that illustrated on the right in 
Fig.~\ref{btosgammavsx} get contributions from the region in which the
loop momentum $l$ is smaller than, say, 1 GeV. The $n!$ factor is the
price we pay for applying perturbation theory in a region in which
perturbation theory does not really apply. 

Of course, this kind of behavior is generic in QCD perturbation theory.
However, the renormalon behavior is particularly bothersome in this
instance. Gardi reported that the bad behavior stems from the use of the
``pole mass'' in the heavy quark propagator. With a reorganized
calculation, he reported that the integration is less sensitive to the
infrared region and the predictive power of the theory can be improved.

\begin{figure}[htb]
\centerline{\includegraphics[width = 8 cm]{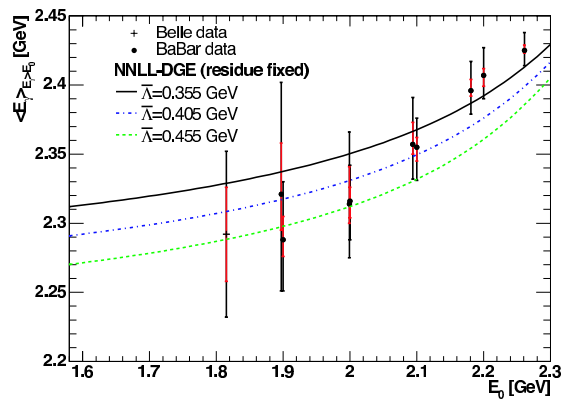} }
\caption{Comparison of experimental results from BaBar and Belle reported
by Walsh \protect\cite{Walsh} to the theory of Andersen and
Gardi.\protect\cite{GardiResult} Here $\bar \Lambda$ is the difference
between the $B$-meson mass and the $b$-quark mass.}
\label{gardi}
\end{figure}

At this conference, J.~Walsh reported new results from BaBar and Belle on 
$d\varGamma/dx$ for this process. Andersen and Gardi \cite{GardiResult}
were able to take this data and compare it to their theoretical results.
They compared theory and experiment for the shape of the spectrum as
represented by
\begin{displaymath}
\frac{1}{\varGamma}
\int_{E_0}^{\infty} dE_\gamma\ E_\gamma \frac{d \varGamma}{d E_\gamma}
\end{displaymath}
as a function of the lower endpoint $E_0$ of the integral. The result is
shown in Fig.~\ref{gardi}. Theory and experiment appear to agree pretty
well.

\section{Pentaquark states}

There was some discussion at the conference of the theoretical
description of the state $\Theta^+(1540)$, which in a quark description
would be a $udud\bar s$ state. At the time of the conference, the
experimental evidence for the existence of this state was mixed. This
state is often discussed as a state consisting of two diquarks and an
antiquark (see the talk of A.~Vainshtein \cite{Vainshtein} on diquarks).
An alternative view, reviewed by M.~Praszalowicz,\cite{Praszalowicz}
derives the state in a chiral quark soliton model. This model is based on
nature being near the chiral limit, in which the $\pi$ and $K$ mesons have
zero mass, and near the limit of having an infinite number, $N_c$, of
colors. Praszalowicz argued that for the $\overline {10}$ representation
of flavor SU(3) to which the $\Theta^+(1540)$ would belong, we are far
from the $N_c \to \infty$ limit. Thus QCD theory does not require the
existence of this state.  That's good, because the existence of this
state is looking more doubtful: on the day this conference ended, the
COMPASS experiment reported its non-observation of the
$\Theta^+(1540)$.\cite{COMPASS}

\section{Lattice studies}

I have mentioned already lattice studies of the matrix elements for weak
interaction decays. We heard two other kinds of lattice studies. One
concerned the gluon propagator, the other was about one of the signals
for the quark gluon plasma in heavy ion collisions. 

\begin{figure}[htb]
\centerline{\includegraphics[width = 8 cm]{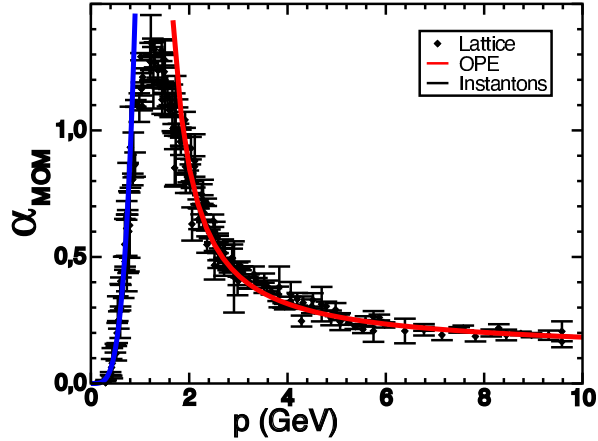} }
\caption{Effective coupling as a function of momentum, defined as
$g^2/4\pi$ where $g$ is defined as a ratio of three and two point Green
functions measured in lattice gauge theory with no dynamical
quarks.\protect\cite{deSoto} The lattice results are compared to an
operator product expansion expectation and to an instanton liquid model.}
\label{desoto}
\end{figure}

Over the years, there have been quite a lot of studies in QCD of gluon
two-point functions and three-point functions, and more generally
$n$-point functions for quarks and gluons. Such studies begin with
perturbation theory since they are based on the Schwinger-Dyson equations,
but they go beyond any fixed order of perturbation theory by solving the
equations within some approximation scheme. Since the $n$-point functions
are gauge dependent, such studies must pick a gauge. F.~de Soto Borrero
reported on studies of the gluon two- and three-point function in QCD
without dynamical quarks,\cite{deSoto} based not on the Schwinger-Dyson
equations but instead on a direct simulation in lattice gauge theory. One
interesting way to express the results is to plot the three point
function divided by the cube of the two point function, all of these
suitably renormalized. That gives the one-particle-irreducible three-point
function and thus an effective coupling that is a function of the
momentum $p$ on each leg of the graph. I should caution that the
effective coupling thus defined is a convenient object to use in
discussing the theory, but is not directly observable in nature. In
Fig.~\ref{desoto}, I display the result, showing that the effective
squared coupling $\alpha_{\rm MOM}$ thus defined increases as the momentum
$p$ decrease, then reaches a maximum and decreases. The nice thing is
that one can understand the large $p$ behavior using the operator product
expansion and the small $p$ behavior using a picture involving the
classical solutions of the (Euclidean) equations of motion known as
instantons.

\begin{figure}[htb]
\centerline{\includegraphics[width = 8 cm]{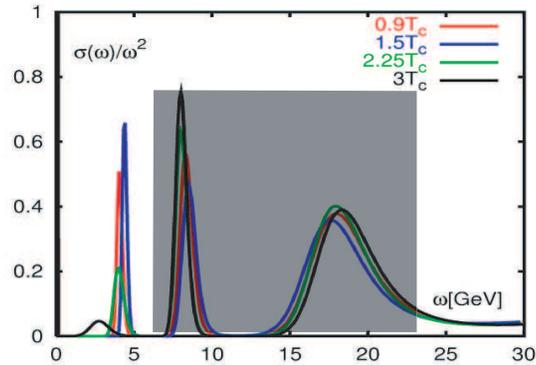} }
\caption{Spectral density in hot QCD as a function of energy $\omega$,
as measured on the lattice.\protect\cite{Petreczky} The data in the
shaded box represents lattice artifacts and should be ignored. }
\label{Jmelting}
\end{figure}

One signal for the quark-gluon plasma that one hopes to
see in heavy ion collisions is the melting of $J/\Psi$ states. At
relatively low temperatures, hadronic matter allows the existence of a 
$J/\Psi$ resonance. But at higher temperatures in the plasma phase, the
$J/\Psi$ can no longer maintain itself. R.~Petreczky presented lattice
results that test this prediction. The results are shown in
Fig.~\ref{Jmelting}. We see that the resonance is present at low
temperatures but for temperatures well above the plasma phase transition
temperature $T_c$, it has melted away. The strength of the resonance has
substantially diminished at $2.25\ T_c \approx 390\ {\rm MeV}$.

\section{Jet Physics}

A number of talks at this conference dealt with the physics of jets. I
will touch on a few topics that seemed to me to be of particular interest.

\begin{figure}[htb]
\centerline{\includegraphics[width = 10 cm]{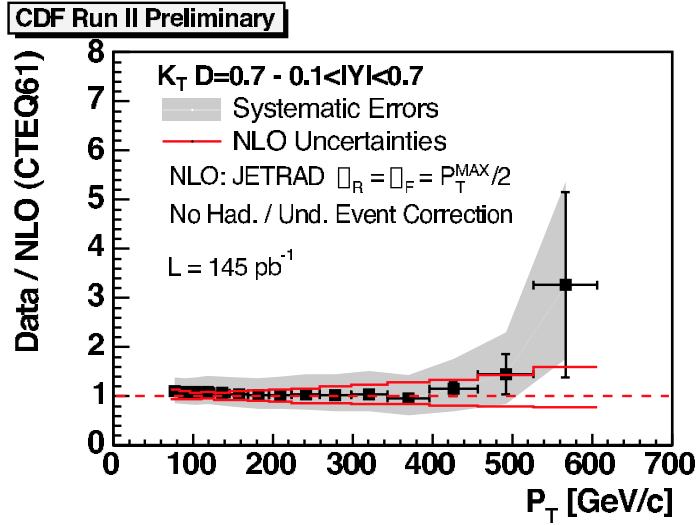} }
\caption{One jet inclusive jet cross section versus $P_T$ of the jet for
jets defined with the $k_T$-algorithm,\protect\cite{Kupco} as measured by
CDF. The graph shows the data divided by the NLO theory prediction, with
experimental systematic errors and estimated theory errors indicated.}
\label{ktjets}
\end{figure}

In Run I at the Fermilab Tevatron, jet cross sections were typically
measured using a cone algorithm. The idea is that a jet consists of
particles whose momentum vectors lie in a cone centered on a jet axis.
This sounds simple. However, it is not so simple when one considers what
to do with overlapping cones. For Run II, the cone algorithm has been
improved with respect to sensitivity to effects from soft particles, but
the complications remain. An alternative is the $k_T$-algorithm, which is
modelled on that used in electron-positron annihilation. This algorithm
is simple enough to state completely in a short paragraph. The idea is
that one successively combines ``nearby'' subjets, thus capturing the
characteristic of QCD that there are jets within jets over a wide range
of transverse momentum scales. At Run II at the Tevatron, CDF and D0 have
been experimenting with the use of this algorithm. A.~Kupco presented data
from CDF showing that the $k_T$-algorithm works well in a practical
sense.\cite{Kupco} The data is displayed in Fig.~\ref{ktjets}.

\begin{figure}[htb]
\centerline{\includegraphics[width = 8 cm]{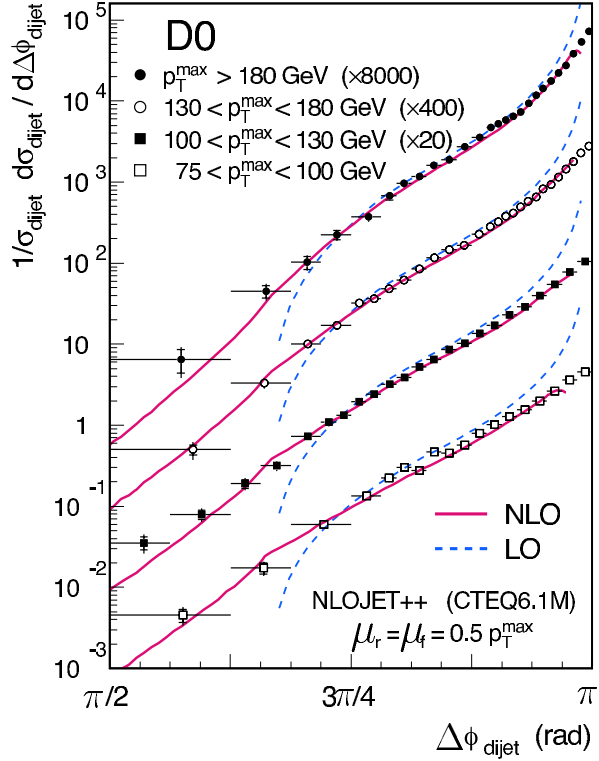} }
\caption{Two jet azimuthal angle correlation \protect\cite{Kupco} as
measured by D0. The distribution is plotted against the azimuthal angle
$\Delta \phi$ between the two jets for four different bins of the
transverse momentum of the jet with the larger transverse momentum. For
$\Delta \phi \ne \pi$ there is at least one more jet that recoils against
the observed jet, so that the Born level for the theory is $\alpha_s^3$.
The data is compared to LO and NLO theoretical predictions.}
\label{deltaphi}
\end{figure}

It has recently become possible to calculate three-jet quantities in
hadron collisions at next-to-leading order. An interesting example of
this is the correlation in azimuthal angle $\Delta\phi$ between two
observed jets (out of three or more jets in the event).  A.~Kupco
\cite{Kupco} showed results from D0 on this quantity. The NLO theory does
well in predicting the result over a wide angular range as seen in
Fig.~\ref{deltaphi}. (In the region close to $\Delta\phi = \pi$ we are
close to the two jet region, so that fixed order perturbation is not
expected to work.) The NLO theory is evidently significantly better than
the LO three-jet theory, which inevitably fails as one gets near
$\Delta\phi = \pi/2$ because the two jets with the largest transverse
momenta cannot have $\Delta\phi = \pi/2$ unless they recoil against at
least two other jets. A separate graph in this talk illustrated that the
theoretical prediction from the Pythia Monte Carlo event generator is
sensitive to the available tuning parameters in the $\Delta\phi \sim
\pi/2$ region. Presumably Pythia would have done better if it matched to
the exact tree level four-jet matrix element. 

Jet physics took a major step forward in the 1980s with the analysis of
three-jet events at the PETRA accelerator at DESY. The data is still
useful, as illustrated in the talk of S.~Kluth.\cite{Kluth} He and
collaborators used data from the Jade experiment at PETRA to extract the
four jet rate at several values of $\sqrt s$ using the Durham (or $k_T$)
jet algorithm. It is the progress of QCD theory that has made this
difficult job worthwhile: neither the Durham algorithm nor a calculation
of four jet rates at next-to-leading order were available at the time of
the Jade experiment. The results provide a new way to extract
$\alpha_s(\sqrt s)$ from the Jade data for $\sqrt s$ in the PETRA energy
range.

\section{Perturbative calculations}

In the prior sections we have seen examples of the progress in QCD theory
over the years, as in the comparison of four-jet data from the Jade
experiment in the early 1980s to NLO four jet calculations that have only
been accomplished in the past few years. Continuing progress in theory
was reported at this conference.

For a ``high value'' cross section like Higgs production via gluon-gluon
fusion, it has been on everyone's wish list to have a
next-to-next-to-leading order calculation. Furthermore, one would like to
be able to calculate the cross section for an arbitrary infrared safe
observable involving the Higgs boson and other partons created in the
process. This is a very difficult problem, and in my opinion the best way
to do this ultimately will be to have a system of subtractions that take
care of all the singularities in the partonic matrix element -- based on
the general singularity structure of QCD. However, progress in this
direction has been slow. C.~Anastasiou discussed a program that does just
this task based not on knowing the singularity structure of QCD but on
letting a computer find the singularities.\cite{Anastasiou} For three
final state partons, one maps the momenta $\{p_1,p_2,p_3\}$ constrained
to have the sum of the transverse momenta vanish into seven variables 
$\{x_1, x_2, x_3, x_4, x_5, x_6, x_7\}$ in a seven dimensional hypercube.
The mapping should be such that the singularities are at the edges and
faces of the hypercube. Then the computer is asked to find the
singularities and make the appropriate subtractions. I can illustrate the
idea in a simple fashion if I use only two variables $x$. Consider an
integral of the form
\begin{equation}
I= \int_0^1 dx_1\, x_1^\epsilon \int_0^1 dx_2 \,\frac{f(x_1,x_2)}{x_1},
\end{equation}
where the factor $x_1^\epsilon$ mimics the result of using dimensional
regularization and $f(x_1,x_2)$ is a smooth function that could be quite
complicated. We would like to separate this into a pole term proportional
to $1/\epsilon$ and a term that is finite as $\epsilon \to 0$. To this
end, we ask our computer to notice the singularity at $x_1 \to 0$ and
write the integral as
\begin{equation}
\begin{split}
I ={}&\int_0^1 dx_1\, x_1^\epsilon \int_0^1 dx_2 
\left\{\frac{f(0,x_2)}{x_1}  +\frac{f(x_1,x_2)-
f(0,x_2)}{x_1}
\right\}
\\
\sim{}&
{\frac{1}{\epsilon}} 
\int_0^1 dx_2 \,f(0,x_2)
+
\int_0^1 dx_1 \int_0^1 dx_2 \
\frac{f(x_1,x_2) - f(0,x_2)}{x_1}
.
\end{split}
\end{equation}
The first term is the pole term, the second is the finite term. Both
integrals can be computed by numerical integration. In real life, the
situation is much more complicated, but it appears from the results
presented in \cite{Anastasiou} that this is a practical way to proceed.

By going from NLO calculations to NNLO calculations, one reduces the
estimated theory error for the prediction of a certain class of cross
sections. There is also another needed direction for improvement of
calculations.

Conventional NLO calculations do not do well at predicting the probability
to emit soft particles or at describing the inner structure of jets. That
is why one restricts their use to the prediction of those cross sections
that are ``infrared safe'' in the sense that they are not sensitive to
soft particles in the final state or to how jets are divided into
subjets. For instance, suppose that one were to take a program that gives
the cross section for $p +  \bar p \to W + {\it jet} + X$ at
NLO. Now imagine using the program to predict the distribution of masses
for the jet. The result would be a certain number of events with jet
masses near 10 GeV, more events with jet masses near 5 GeV, many more
with jet masses near 1 GeV, and yet more with yet smaller jet masses. This
all adds up to an infinite number of events, but the situation is saved
by having an infinitely negative number of events with jet mass zero.
Evidently, this is not a satisfactory description of nature. Nor is it
satisfactory that the jets are made of one or two partons rather than
many hadrons.

Clearly it would be better to let the partons fragment to form parton
showers and then let the parton showers form hadrons in the style of
event generator Monte Carlo programs like Pythia and Herwig. It is,
however, not so easy to do this while maintaining the next-to-leading
order accuracy of the calculation for the infrared safe quantities that
the NLO program was designed to get right.

There has been some progress in this area in recent
years,\cite{NLOShowers} but the presently existing programs are either
tied to a specific Monte Carlo event generator or a limited class of
processes and are not specified in a simple general algorithm that
authors of NLO calculations could easily use. 

At this conference, Z.~Nagy \cite{Nagy} presented an algorithm for
extending an NLO calculation in a fashion that would allow partonic
events from the modified NLO program to be fed to a Monte Carlo event
generator for showering and hadronization, on the condition that the
showering be started for each event with suitable initial conditions.
The algorithm is presented for jet production in electron-positron
annihilation, but it should be possible to extend it to lepton-hadron and
hadron-hadron collisions. The idea is to base the algorithm on the
Catani-Seymour dipole subtraction scheme \cite{CataniSeymour} that is
widely used in constructing NLO programs. In addition, the algorithm
makes use of the $k_T$ jet-matching scheme \cite{CKKW} for switching
descriptions among, for example, $p +  \bar p \to W + X$, $p +  \bar p
\to W + 1\ {\it jet} + X$, and $p +  \bar p \to W + 2\ {\it jets} + X$.
This is so far an algorithm but not yet code. Perhaps we will see a
working program at the XLI Rencontre de Moriond.

\section{QCD in the high energy/small $x$ limit}

At this conference there was quite a lot of discussion of how QCD behaves
when seen with experiments that, in one view, probe the gluon
distribution at small momentum fraction $x$. Relevant experiments include
deeply inelastic scattering at small Bjorken $x$ and heavy ion collisions
at RHIC.

An outsider like me can understand at least part of this discussion with
the aid of a few of simple observations. First, the whole subject hinges
on how systems with vastly different rapidities can interact. For
instance, if one is interested in the total cross section for scattering
two nuclei at high energy, the rapidity difference $\Delta y = \log
(s/M^2)$ is large. In deeply inelastic scattering, the rapidity
difference between the virtual photon and the proton is $\Delta y =
\log(Q/m) + \log (1/x)$ and is large if $x$ is small (even if $Q/m$
is not so large). Second, systems with a large rapidity difference can
communicate easily by exchanging gluons, so that the whole subject hinges
on gluons. Third, experiment shows that lots of gluons are to be found in
the proton carrying a small fraction $x$ of the proton's momentum. Thus
the small $x$ gluons appear to be densely packed inside the transverse
size of a proton or nucleus. With respect to a probe that might measure
this gluon distribution by scattering from it, the hadron looks black.
This suggests a system that is purely non-perturbative and thus resistant
to theoretical analysis. However, it is useful to think of our probe as a
fast moving color dipole. For instance, in deeply inelastic scattering,
the virtual photon creates a quark-antiquark pair with net color zero --
that is a color dipole. This leads to the fifth observation: If the dipole
is small enough, the dense gluonic system is transparent to the
dipole. Thus there is some size $1/Q_s(x)$ for which the gluonic system
just starts to scatter the dipole. At small $x$, the density of gluons
is so high $1/Q_s(x)$ is small. That is, the ``saturation scale''
$Q_s(x)$ is big. This means that there is naturally a big transverse
momentum scale in the problem and perturbation theory can be helpful.
This analysis does not carry us very far into the subtleties of the
theory, but it is perhaps useful in getting us started.

\begin{figure}[htb]
\centerline{
\includegraphics[width = 7 cm]{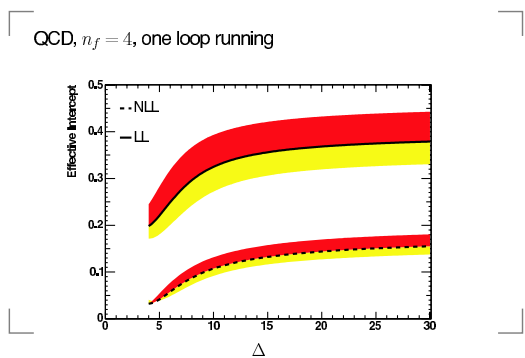}}
\caption{Behavior of the effective pomeron intercept according to the
BFKL equation as a function of rapidity $\Delta$ from the talk of
J.~Andersen.\protect\cite{Andersen} Solutions are shown for the lowest
order equation and for the next-to-leading order equation solved by an
iterative numerical method. Note that the effective intercept at
next-to-leading order is fairly small and approximately constant as
$\Delta$ increases.}
\label{BFKLsoln}
\end{figure}

One approach to the high energy limit of QCD is the much studied BFKL
equation. There is a leading order version of this equation and, more
recently, next-to-leading order corrections. However, it has appeared
that the general approach is unstable, with the corrections leading to
big changes in the behavior of the solutions. At this conference,
L.~Schoeffel \cite{Schoeffel} discussed ways of modifying the
equation so as to avoid some of the problems and showed results for the
solutions of the resulting equation. J.~Andersen \cite{Andersen} took a
different approach, arguing that the widely used Mellin transform method
of solving the BFKL equation leads to problems, so that one should instead
use an iterative numerical approach. If one defines an effective pomeron
intercept  $\alpha_{I\kern - 5 pt P}$ by $x g(x ) \sim x^{-\alpha_{I\kern
- 5 pt P}}$, then Andersen finds a sensible behavior for 
 $\alpha_{I\kern - 5 pt P}$ as a function of rapidity, as shown in
Fig.!\ref{BFKLsoln}.

Another approach to understanding the high energy limit of QCD is to go
beyond the BFKL equation. There were separate talks on this by
E.~Iancu,\cite{Iancu} by D.~Triantafyllopoulos \cite{Triantafyllopoulos}
and by G.~Soyez.\cite{Soyez}  Here, I follow mostly Iancu and
Triantafyllopoulos. One considers the interaction of a color dipole with
a hadron to be created by the exchange of $n$ gluon ladders, or
``pomerons.'' Letting
$T_n$ be contribution from for $n$ pomerons, one writes an equation for
the variation of $T_n$ with the rapidity $y$ of the dipole relative to the
hadron,
\begin{equation}
dT_n/dy =  K_{\rm BFKL}\, T_n
+ K_{\rm split}\, T_{n-1}
+ K_{\rm merge}\, T_{n+1}\ .
\end{equation}
Here the $T_n$ are functions of transverse position arguments and the
various $K$ are kernels in an integral equation. This general scheme
incorporates the linear evolution of a gluon ladder, the splitting of one
ladder to make two, and the joining of two gluon ladders to make one. The
first term, for $n=1$ is the leading order BFKL equation. 
Triantafyllopoulos argued that the splitting and merging terms are more
important than the next-to-leading order corrections to the BFKL equation.
Iancu argued that one has some hope of making progress with the
splitting-joining picture, even though it is complicated, since the
problems are related to problems that have been studied in statistical
physics.

\section{Automating the scientific method}

Part of the purpose of this talk was to provide some
assessment of where we stand with respect to the problems and
opportunities facing QCD theory. Of particular interest is the upcoming
beginning of the LHC era. One anticipates that what is seen may not be
in agreement with the Standard Model. Since the incoming particles and
many of the outgoing particles in an LHC event carry the strong
interactions, one is going to have to incorporate QCD theory into the
interpretation of whatever signals may be found. Are we ready?

B.~Knutsen \cite{Knutsen} presented computer tools that can help in the
interpretation of the data. First, there are tools for finding
discrepancies between the data and Standard Model predictions. Then,
there are tools for testing hypotheses against the data. Finally, there
is a tool for generating hypotheses and testing them. One might call this
last step automating theorists.

I am a bit skeptical about this last part. Of course, it is not the
theorists who construct computer programs for doing calculations who are
in danger of being replaced by automatons: their programs would just be
added to the array of programs that does hypothesis testing. Rather, it
is the model building theorists whom we perhaps don't need. However, my
guess is that we will be faced with a more difficult problem than simply
filling in the blanks for the Minimal Supersymmetric Standard Model, so
that some clever ideas from people like those who built the MSSM will be
required.

Overall, it seems to me that we ought to automate whatever we can.
However, I was struck by Knutsen's comment that high energy physics is
behind astrophysics, heavy ion physics, and biology with respect to
automation. Accordingly, I would like to present a real example from
current developmental biology that provides an interesting comparison to
high energy physics.

\begin{figure}[htb]
\centerline{
\includegraphics[width = 3 cm]{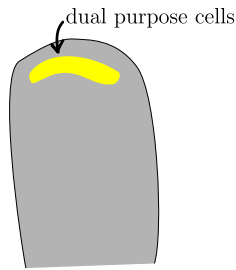}
}
\caption{Sketch of region dual purpose cells in a zebrafish embryo.}
\label{fish1}
\end{figure}

\begin{figure}[htb]
\centerline{
\includegraphics[width = 6 cm]{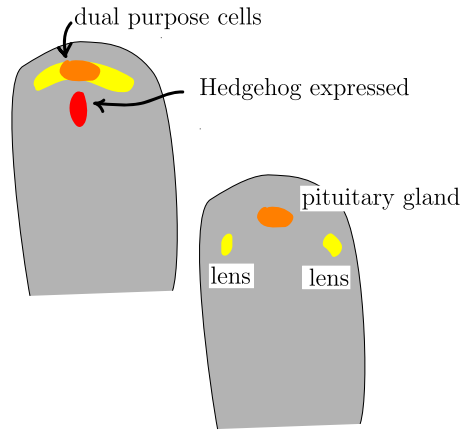}
}
\caption{Sketch illustrating the role of Hedgehog signalling protein in
determining the fate of the dual purpose cells in
Fig.~\protect\ref{fish1}.}
\label{fish2}
\end{figure}

The example concerns the zebrafish, a small tropical fish that is easily
raised in tanks. This fish has the wonderful feature that its
embryos are quite transparent, so that the parts of an embryo can be
studied under a microscope as the embryo develops. Study of how the
embryo develops can help us figure out the mechanisms for development. One
finds \cite{fish} that in the early embryo there  is a band of cells that
are able to be either pituitary gland cells or else cells in the lens of
the fish's eye (Fig.~\ref{fish1}). How, then, do these cells decide
what to do? Experiments \cite{fish} show that there are some nearby cells
that express a signalling molecule called Hedgehog. The dual purpose
cells that get a strong Hedgehog signal become pituitary gland cells,
while those that get a weaker signal become lens cells
(Fig.~\ref{fish2}). One can test this by using a mutant fish that does
not produce Hedgehog. This amounts to changing the DNA of the starting
fish egg.

\begin{figure}[htb]
\centerline{
\includegraphics[width = 6 cm]{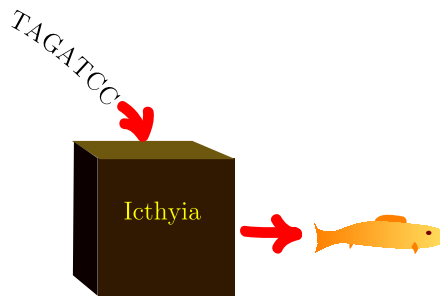}
}
\caption{How to automate developmental biology. Build a computer program
that takes a DNA sequence as input and models the development of a fish.}
\label{icthyia}
\end{figure}

The experiments in biology that lead to such discoveries are difficult.
Things would be easier if biologists could automate the process by
producing a computer program, which might be called Icthia, that could
simulate fish development (Fig.~\ref{icthyia}). The input would be the
fish genome. Then the program would simulate the complete development of
the fish, including the signalling pathways just discussed. In the end, we
would have a model final state fish to compare to an experimental fish.

\begin{figure}[htb]
\centerline{
\includegraphics[width = 6 cm]{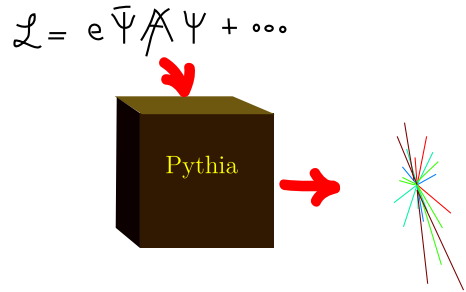}
}
\caption{How particle physics has already been automated. Pythia takes
the lagrangian as input and models the development of $pp$ collision
events. Herwig does this too. (Of course, there are limitations, for
instance with respect to the range of input lagrangians that are
programmed and with respect to the accuracy of the output.)}
\label{pythia}
\end{figure}

Of course, there is no such program. Biology is too hard for that. But
physics is easier. We do have programs, for instance Pythia, that take
the lagrangian of the Standard Model or various extensions as input and
simulate the entire development of high energy physics events, from an
initial state shower to a hard interaction such as, say, a quark and
antiquark producing a squark and antisquark, followed by parton showering
and decays, followed by hadronization, producing final states of events
that can be compared to experimental events (Fig.~\ref{pythia}).

We thus have, I submit, theoretical tools for the analysis of the upcoming
LHC experiments that are extraordinarily powerful. The event generator
Monte Carlo programs just described have a very wide scope. There are
also well automated tools for creating tree level cross sections
for an even wider variety of partonic processes. Where more accuracy is
needed for suitably inclusive cross sections, we have NLO programs and
even some NNLO programs. These tools have been developed over the last
couple of decades and are now, I think, adequate to the task at hand.
Furthermore, these theory tools are currently being improved. It was a
pleasure to hear about some of the progress at this conference. The more
improvements we have, the better off we will be in trying to understand
what we find at the LHC.

\end{document}